# Photonic Floquet media with a complex time-periodic permittivity

Neng Wang, Zhao-Qing Zhang, and C. T. Chan*

*Department of physics, The Hong Kong University of Science and Technology, Clear Water Bay, Hong Kong, China*

Corresponding author: phchan@ust.hk.

## Abstract

We study the formation of exceptional point (EP) phenomena in a photonic medium with a complex time-periodic permittivity, i.e., $\varepsilon(t)=\varepsilon_o+\varepsilon_r\sin(\Omega t)$. We formulate the Maxwell's equations in a form of first-order non-Hermitian Floquet Hamiltonian matrix and solve it analytically for the Floquet band structures. In the case when $\varepsilon_r$ is real, to the first order in $\varepsilon_r$, the band structures show a phase transition from an exact phase with real quasienergies to a broken phase with complex quasienergies inside a region of wave vector space, the so-called k-gap. We show that the two EPs at the upper and lower edges of the k-gap have opposite chiralities in the stroboscopic sense. Thus, by picking up the mode with a positive imaginary quasienergy, the wave propagation inside the k-gap can grow exponentially. In three dimensions, such pairs of EPs span two concentric spherical surfaces in the $\vec{k}$ space and repeat themselves periodically in the quasienergy space with $\Omega$ as the period. However, in the case when $\varepsilon_r$ is purely imaginary, the k-gap disappears and gaps in the quasienergy space are opened. Our analytical results agree well with the finite-difference time-domain (FDTD) simulations. To the second order in $\varepsilon_r$, additional EP pairs are found for both the cases of real and imaginary $\varepsilon_r$. We also extend our theory to the case where the permittivity is simultaneously modulated in both space and time, i.e., $\varepsilon(x,t)=\varepsilon_o+\varepsilon_r\sin(\Omega t-\beta x+\phi)$. For a real $\varepsilon_r$, we find that EPs also exist when the modulation speed $c_m=\Omega/\beta$ is faster than the speed of electromagnetic wave inside the medium.

## I. Introduction

Time-Floquet or periodically driven systems are the systems whose parameters are periodic in time. As the time modulation can represent a rich and versatile resource that is used to achieve many novel phenomena, time-Floquet systems have attracted great attention recently. In quantum systems, researchers have reported that the topological spectra can be achieved when a proper time perturbation is introduced to a system which is topologically trivial in the static case [1-3], and the topological state is called the Floquet topological state. Later, these Floquet topological states were observed in photonic [4-6] and solid-state [7] experiments, which motivated the detailed study of topological phenomena in periodically driven systems [8-12]. Time-Floquet classical systems, such as the LC circuits with time-reactive elements [13-17] and the dynamic mediums with time-periodic permittivity [18-24], have been investigated. For a long time, researchers have focused their study on the amplification and non-reciprocal behaviors in these time-Floquet classical systems [25-29] and made great achievements. Recently, there is also a surge of interests in investigating the systems that combine both the non-Hermiticity and time-periodic modulations [30-36], leading to some novel phenomena that cannot be found in static systems, such as the exotic parity-time (PT) transitions [34, 35] and non-reciprocal gain without gain materials [36].

In this paper, we study a typical time-Floquet photonic system in which the permittivity of the medium has the form $\varepsilon(t)=\varepsilon_o+\varepsilon_r\sin(\Omega t+\phi)$, where $\varepsilon_r$ gives the strength of time modulation and can be a complex number in general, $\Omega$ is the modulation frequency and $\phi$ is an arbitrary initial phase which sets the origin of the time. When $\varepsilon_r$ is real, dispersion relations as well as transmission and reflection properties of such dynamic media have been studied previously by solving a second-order time-dependent scalar wave equation for the electric field [21,22]. Here, we are interested in the exceptional point (EP) phenomena in such systems. Using the method of Floquet matrix [37], we formulate the Maxwell's equations in the form of a first-order non-Hermitian Floquet Hamiltonian matrix and we solve for the photonic Floquet band structures as well as the Floquet states. To the first order in $\varepsilon_r$, we show explicitly that the existence of a gap in the wave vector space (the so-called k-gap) is an EP phenomenon. At both the upper and lower edges of the k-gap, two real quasienergies coalesce and form a pair of EPs which are always in opposite chiralities in the stroboscopic sense. The region between these two EPs is a broken phase, in which the quasienergies form complex conjugate pairs so that the Floquet states can decay or grow. The amplification and damping of waves in this region are induced by the time modulation in the absence of gain and lossy materials [34-36]. By picking up the mode with a positive imaginary quasienergy, the wave propagation inside the k-gap always grows exponentially with a maximal growth rate near the center of the k-gap. This is numerically verified by using the finite-difference time-domain (FDTD) simulations. In three dimensions, all these EP pairs in different directions in the $\vec{k}$ space form two concentric spherical surfaces of EPs with a broken phase in between. These two concentric surfaces repeat themselves periodically in the quasienergy space with $\Omega$ as the period. When $\varepsilon_r$ is purely imaginary, we find quasienergy gaps instead of a k-gap. The opening of quasienergy gaps in this case is analogous to the case of graphene where a gap near a Dirac point can be opened when the onsite energies of the electrons at two atoms in the unit cell are different. However, when the difference in the onsite energies is imaginary, the static Dirac Hamiltonian becomes non-Hermitian and the Dirac point splits into two EPs with opposite chiralities. The latter case corresponds to the case of real $\varepsilon_r$ discussed earlier. There is no EP when $\varepsilon_r$ is a complex number. Since the quasienergies still form complex conjugate pairs, the medium will support field amplification and damping. To second order in $\varepsilon_r$, we find additional pairs of EPs for both pure real and pure imaginary $\varepsilon_r$.

Our theory can also be extended to the more complicated case where the permittivity is simultaneously modulated in both space and time, which has the form $\varepsilon(x,t) = \varepsilon_o + \varepsilon_r \sin(\Omega t - \beta x + \phi)$ with $\beta$ being the spatial modulation frequency. Using the new variable $u = t - \beta x / \Omega$ instead of $t$, we analytically obtained the Floquet Hamiltonian and used it to calculate the Floquet bands. When $\varepsilon_r$ is real, we also find EPs in the Floquet bands when the modulation speed $c_m = \Omega / \beta$ is faster than the speed of electromagnetic wave inside the medium.

Temporally modulating the permittivity in experiments is not easy. Nevertheless, various techniques such as electro-optic, thermal-optic and plasma dispersion effects have been proposed for realizing time modulation [37-44]. Owing to the growing interests in the time Floquet photonic system, sustainable time modulation on the permittivity with very high modulation frequency can be feasible in the near future [24].

## II. Formulation of the Floquet Hamiltonian

We consider the electromagnetic wave propagation inside a dynamic medium where the permittivity is time periodic. Without loss of generality, we consider a wave propagating along the *x* direction with the electric field and magnetic field in the *y* and *z* directions, respectively. For the sake of mathematical simplicity, we set both the permittivity and permeability in vacuum as unity and assume that the static relative permittivity $\varepsilon_o > 0$. By using the Maxwell equations

$$\begin{aligned}\nabla\times\mathbf{E} &= -\frac{\partial}{\partial t}\mathbf{B} = -\frac{\partial\mathbf{H}}{\partial t},\\ \nabla\times\mathbf{H} &= \frac{\partial}{\partial t}\mathbf{D} = [\varepsilon_o+\varepsilon_r\sin(\Omega\mathrm{t}+\phi)]\frac{\partial\mathbf{E}}{\partial t} + \Omega\varepsilon_r\cos(\Omega\mathrm{t}+\phi)\mathbf{E},\end{aligned} \tag{1}$$

we obtain the following wave equations for the electromagnetic fields:

$$\begin{aligned}&-[\varepsilon_o+\varepsilon_r\sin(\Omega\mathrm{t}+\phi)]\frac{\partial^2 E_y}{\partial t^2} - 2\Omega\cos(\Omega\mathrm{t}+\phi)\varepsilon_r\frac{\partial E_y}{\partial t} + \Omega^2\varepsilon_r\sin(\Omega\mathrm{t}+\phi)E_y = -\frac{\partial^2 E_y}{\partial x^2} = K^2 E_y,\\ &-[\varepsilon_o+\varepsilon_r\sin(\Omega\mathrm{t}+\phi)]\frac{\partial^2 H_z}{\partial t^2} - \Omega\cos(\Omega\mathrm{t}+\phi)\varepsilon_r\frac{\partial H_z}{\partial t} = -\frac{\partial^2 H_z}{\partial x^2} = K^2 H_z,\end{aligned} \tag{2}$$

where $\mathbf{K} = K\hat{x}$ is the wave vector. To obtain the Floquet Hamiltonian, we followed Ref. [34] to reduce the differential order by employing the Liouvillian formulation and rewrite Eq. (2) as a two-component time-dependent Schrodinger-like equation for the magnetic field and its time derivative, i.e.,

$$i\frac{\partial}{\partial t}\begin{pmatrix} H_z \\ \dot{H}_z \end{pmatrix} = \hat{\mathrm{H}}_{eff}\begin{pmatrix} H_z \\ \dot{H}_z \end{pmatrix} = \begin{pmatrix} 0 & i \\ -iB & -iA \end{pmatrix}\begin{pmatrix} H_z \\ \dot{H}_z \end{pmatrix}, \tag{3}$$

where $\dot{H}_z = \partial H_z / \partial t$ and

$$A = \frac{\Omega\varepsilon_r\cos(\Omega\mathrm{t}+\phi)}{\varepsilon_o+\varepsilon_r\sin(\Omega\mathrm{t}+\phi)}, \qquad B = \frac{K^2}{\varepsilon_o+\varepsilon_r\sin(\Omega\mathrm{t}+\phi)}. \tag{4}$$

In the absence of time modulation, i.e., $\varepsilon_r = 0$, $\hat{\mathrm{H}}_{eff}$ reduces to the time-independent non-Hermitian form:

$$\hat{\mathrm{H}}_{eff,0} = \begin{pmatrix} 0 & i \\ -i\dfrac{K^2}{\varepsilon_o} & 0 \end{pmatrix}. \tag{5}$$

The Floquet theorem gives the so-called Floquet-state solution of Eq. (3) for a specific wave vector $K$ as

$$\begin{pmatrix} H_z \\ \dot{H}_z \end{pmatrix} = e^{-iQ(K)t}\vec{\Phi}(x,t) \tag{6}$$

where $Q$ is the Floquet characteristic exponent or the so-called quasienergy, which is $K$-dependent, and $\vec{\Phi}$ is the so-called Floquet mode which is periodic in time obeying $\vec{\Phi}(x,t) = \vec{\Phi}(x,t+2\pi/\Omega)$. Substituting Eq. (6) into Eq. (3), one obtains the time-dependent eigenvalue equation for the quasienergy $Q$ as

$$(\hat{\mathrm{H}}_{eff} - i\frac{\partial}{\partial t})\vec{\Phi} = Q\vec{\Phi}. \tag{7}$$

From Eqs. (6) and (7), we can see that if $Q$ is an eigenvalue of Eq. (7) corresponding to the eigenstate $\vec{\Phi}$, then $Q + n\Omega$ with $n$ being an arbitrary integer number is also an eigenvalue of Eq. (7) corresponding to the eigenstate $e^{in\Omega t}\vec{\Phi}$. Similar to the Bloch wave vector of a photonic crystal with discrete translational symmetry, the quasienergy can be thought of as a periodic variable defined on a quasienergy Brillouin zone $0 \le Q \le \Omega$.

In the following, we will use the Floquet matrix method [45] to obtain the band dispersions $Q(K)$. The main idea of the method is to use the eigenvectors of the time-independent effective Hamiltonian $\hat{\mathrm{H}}_{eff,0}$ as the basis to expand the Floquet mode $\vec{\Phi}$ and then transfer the time-dependent eigenvalue problem Eq. (7) into a time-independent one. Since $\hat{\mathrm{H}}_{eff,0}$ is pseudo-Hermitian [46], we can apply a similarity transformation on $\hat{\mathrm{H}}_{eff,0}$ to make the time-independent effective Hamiltonian Hermitian. Here we introduce a time-independent matrix [47]

$$\hat{R} = \begin{pmatrix} \sqrt{C} & i \\ \sqrt{C} & -i \end{pmatrix}, \qquad C = \frac{K^2}{\varepsilon_o}, \tag{8}$$

and the time-dependent effective Hamiltonian is then transformed to

$$\tilde{\mathrm{H}}(t) = \hat{R} \cdot \hat{\mathrm{H}}_{eff}(t) \cdot \hat{R}^{-1} = \begin{pmatrix} \dfrac{B - iA\sqrt{C} + C}{2\sqrt{C}} & \dfrac{B + iA\sqrt{C} - C}{2\sqrt{C}} \\ \dfrac{-B + iA\sqrt{C} + C}{2\sqrt{C}} & \dfrac{-B - iA\sqrt{C} - C}{2\sqrt{C}} \end{pmatrix}. \tag{9}$$

The Floquet modes are then transformed according to $\vec{\Phi}' = \hat{R} \cdot \vec{\Phi}$ . In the absence of time modulation ($\varepsilon_r = 0$), the time-independent part of the effective Hamiltonian becomes

$$\tilde{\mathrm{H}}_0 = \hat{R} \cdot \hat{\mathrm{H}}_{eff,0} \cdot \hat{R}^{-1} = \frac{K}{\sqrt{\varepsilon_o}} \begin{pmatrix} 1 & 0 \\ 0 & -1 \end{pmatrix}, \tag{10}$$

which is Hermitian with the eigenvectors being $\vec{\varphi}_1 = (1,0)^T e^{iKx}$ and $\vec{\varphi}_2 = (0,1)^T e^{-iKx}$, respectively. They represent two linear dispersions; a positive band ($\Omega = cK$) and negative band ($\Omega = -cK$) where $c = 1/\sqrt{\varepsilon_o}$ . Then the new Floquet mode can be expanded in terms of the complete orthonormal set $\{\vec{\varphi}_1, \vec{\varphi}_2\}$ as

$$\vec{\Phi}' = \sum_{\beta=1}^{2} \sum_{n=-\infty}^{\infty} c_\beta^n \vec{\varphi}_\beta e^{in\Omega t}, \tag{11}$$

where $c_\beta^n$ is the expansion coefficients. Substituting Eqs. (11) and (9) into Eq. (7), multiplying $(\vec{\varphi}_\alpha)^\dagger e^{-il\Omega t}$ from the left and integrate after a time-average, we have the time-independent eigenvalue problem for the qusienergy $Q$

$$\sum_{\beta=1}^{2} \sum_{n=-\infty}^{\infty} \langle \vec{\varphi}_\alpha l \,|\, H_F \,|\, \vec{\varphi}_\beta n \rangle c_\beta^n = Q c_\alpha^l, \tag{12}$$

with the Floquet Hamiltonian defined as

$$\langle \alpha, n \,|\, H_F \,|\, \beta, l \rangle = \tilde{H}^{n-l}_{\alpha\beta} + n\Omega\delta_{\alpha\beta}\delta_{nl}, \tag{13}$$

where $-\infty \le n, l \le \infty$ label the block matrix, $\alpha, \beta = 1, 2$ label the components inside the block matrix, and

$$\tilde{H}^{n-l}_{\alpha\beta} = \frac{\Omega}{2\pi} \int_0^{2\pi/\Omega} (\vec{\varphi}_\alpha)^\dagger \cdot \tilde{\mathrm{H}}(t) \cdot \vec{\varphi}_\beta e^{i(n-l)\Omega t} dt, \quad \alpha, \beta \in (1, 2). \tag{14}$$

In the limit of small $|\varepsilon_r / \varepsilon_o|$, we can expand $\tilde{\mathrm{H}}(t)$ using Taylor series. To the first order in $\varepsilon_r / \varepsilon_o$ we find

$$\tilde{\mathrm{H}}(t) = \tilde{\mathrm{H}}_0 + \tilde{\mathrm{H}}_1 e^{i\Omega t} + \tilde{\mathrm{H}}_{-1} e^{-i\Omega t}, \tag{15}$$

where

$$\begin{aligned}
\tilde{\mathrm{H}}_1 &= \tilde{\mathrm{H}}_{1,K} + \tilde{\mathrm{H}}_{1,\Omega} = e^{i\phi} \frac{i\varepsilon_r}{4\varepsilon_o} \frac{K}{\sqrt{\varepsilon_o}} \begin{pmatrix} 1 & 1 \\ -1 & -1 \end{pmatrix} + e^{i\phi} \frac{i\varepsilon_r}{4\varepsilon_o} \Omega \begin{pmatrix} -1 & 1 \\ 1 & -1 \end{pmatrix}, \\
\tilde{\mathrm{H}}_{-1} &= \tilde{\mathrm{H}}_{-1,K} + \tilde{\mathrm{H}}_{-1,\Omega} = e^{-i\phi} \frac{i\varepsilon_r}{4\varepsilon_o} \frac{K}{\sqrt{\varepsilon_o}} \begin{pmatrix} -1 & -1 \\ 1 & 1 \end{pmatrix} + e^{-i\phi} \frac{i\varepsilon_r}{4\varepsilon_o} \Omega \begin{pmatrix} -1 & 1 \\ 1 & -1 \end{pmatrix}.
\end{aligned} \tag{16}$$

From Eq. (13), the Floquet Hamiltonian has the form

$$\tilde{\mathrm{H}}_F = \begin{pmatrix} \ddots & \ddots & 0 & 0 & 0 \\ \ddots & \tilde{\mathrm{H}}_0 + \Omega \overleftrightarrow{\mathrm{I}}_2 & \tilde{\mathrm{H}}_1 & 0 & 0 \\ 0 & \tilde{\mathrm{H}}_{-1} & \tilde{\mathrm{H}}_0 & \tilde{\mathrm{H}}_1 & 0 \\ 0 & 0 & \tilde{\mathrm{H}}_{-1} & \tilde{\mathrm{H}}_0 - \Omega \overleftrightarrow{\mathrm{I}}_2 & \ddots \\ 0 & 0 & 0 & \ddots & \ddots \end{pmatrix}. \tag{17}$$

where $\overleftrightarrow{\mathrm{I}}_2$ is the $2 \times 2$ identity matrix. To the second order in $\varepsilon_r / \varepsilon_o$, we find

$$\tilde{\mathrm{H}}(t) = \tilde{\mathrm{H}}_0 + \tilde{\mathrm{H}}_1 e^{i\Omega t} + \tilde{\mathrm{H}}_{-1} e^{-i\Omega t} + \tilde{\mathrm{H}}_2 e^{2i\Omega t} + \tilde{\mathrm{H}}_{-2} e^{-2i\Omega t}, \tag{18}$$

where

$$\tilde{H}_2 = \tilde{H}_{2,K} + \tilde{H}_{2,\Omega} = e^{2i\phi}\frac{\varepsilon_r^2}{8\varepsilon_o^2}\frac{K}{\sqrt{\varepsilon_o}}\begin{pmatrix} -1 & -1 \\ 1 & 1 \end{pmatrix} + e^{2i\phi}\frac{\varepsilon_r^2}{8\varepsilon_o^2}\Omega\begin{pmatrix} 1 & -1 \\ -1 & 1 \end{pmatrix},$$
$$\tilde{H}_{-2} = \tilde{H}_{-2,K} + \tilde{H}_{-2,\Omega} = e^{-2i\phi}\frac{\varepsilon_r^2}{8\varepsilon_o^2}\frac{K}{\sqrt{\varepsilon_o}}\begin{pmatrix} -1 & -1 \\ 1 & 1 \end{pmatrix} + e^{-2i\phi}\frac{\varepsilon_r^2}{8\varepsilon_o^2}\Omega\begin{pmatrix} -1 & 1 \\ 1 & -1 \end{pmatrix}. \quad (19)$$

And the Floquet Hamiltonian has the form

$$\tilde{H}_F = \begin{pmatrix} \ddots & \ddots & \ddots & 0 & 0 \\ \ddots & \tilde{H}_0 + \Omega\overleftrightarrow{I}_2 & \tilde{H}_1 & \tilde{H}_2 & 0 \\ \ddots & \tilde{H}_{-1} & \tilde{H}_0 & \tilde{H}_1 & \ddots \\ 0 & \tilde{H}_{-2} & \tilde{H}_{-1} & \tilde{H}_0 - \Omega\overleftrightarrow{I}_2 & \ddots \\ 0 & 0 & \ddots & \ddots & \ddots \end{pmatrix}. \quad (20)$$

Similarly, we can generalize the above procedure to any order *n* in $\varepsilon_r / \varepsilon_o$ by finding $\tilde{H}_{\pm n}$ first and then write down the corresponding Floquet Hamiltonian using Eq. (13). Here, we will only consider the cases of $n$ = 1 and 2.

## III. Exceptional points

In the limit of vanishing $|\varepsilon_r / \varepsilon_o|$, Eqs. (16-20) show that the Floquet bands can be obtained by copying the two linear bands of $\tilde{H}_0$ and shifting them up and down in the quasienergy space by integer multiples of $\Omega$, as shown in Fig .1(a). We will call the bands formed by shifting the band of $\tilde{H}_0$ by $n\Omega$ as the *n*th order bands. We note that there are both positive (with positive group velocity) and negative bands (with negative group velocity) for any order *n*. Bands of different orders can cross each other and produce infinite number of diabolic points occurring periodically in the K space , at $K_n = n\sqrt{\varepsilon_o}\Omega / 2$, where *n* is an integer.

When $\varepsilon_r \neq 0$, a positive band and a negative band differing by an order *n* can be coupled by the off-diagonal matrices $\tilde{H}_{\pm n}$ which lifts the degeneracy at the diabolic point and produces a pair of EPs, a quasienergy gap or the mixture of the two, depending on the properties of $\varepsilon_r$. Since the quasienergies in the Floquet bands are periodic with the periodic $\Omega$, we will only discuss quasienergies in the reduced zone $0 \le Q \le \Omega$.

The block matrices $\tilde{\mathrm{H}}_{\pm 1}$ give the coupling between a positive band and a negative band having orders differ by one. To the first order in $\varepsilon_r / \varepsilon_o$, we will only consider the diabolic points located at $Q = \Omega / 2$ and $K = \sqrt{\varepsilon_o}\Omega / 2$. When $\varepsilon_r$ is a real number, $\tilde{\mathrm{H}}_{\pm 1}$ is non-Hermitian as can be seen from Eqs. (16) and (21) below. Two bands in the vicinity of the diabolic points can merge and form two EPs at $K = K_-$ and $K = K_+$, as shown in Fig. 1(b). Quasienergies of the Floquet states inside the broken phase $K_- < K < K_+$ form complex conjugate pairs. To show this explicitly, we consider the following $2 \times 2$ reduced Floquet Hamiltonian involving only a negative band of the first order and a positive band of the zeroth order:

$$\tilde{\mathrm{H}}_{F,red} = \begin{pmatrix} -\frac{K}{\sqrt{\varepsilon_o}} + \Omega & e^{i\phi}\frac{i\varepsilon_r}{4\varepsilon_o}(-\frac{K}{\sqrt{\varepsilon_o}} + \Omega) \\ e^{-i\phi}\frac{i\varepsilon_r}{4\varepsilon_o}(-\frac{K}{\sqrt{\varepsilon_o}} + \Omega) & \frac{K}{\sqrt{\varepsilon_o}} \end{pmatrix}, \tag{21}$$

where the 4 matrix elements are obtained from the corresponding entries in Eq. (17). $\tilde{\mathrm{H}}_{F,red}$ is valid as long as $|\varepsilon_r / \varepsilon_o| \ll 1$. The eigenvalues of $\tilde{\mathrm{H}}_{F,red}$ are given by

$$Q = \frac{\Omega}{2} \pm \frac{1}{2}\sqrt{(\frac{2K}{\sqrt{\varepsilon_o}} - \Omega)^2 - \frac{\varepsilon_r^2}{4\varepsilon_o^2}(\frac{K}{\sqrt{\varepsilon_o}} - \Omega)^2}. \tag{22}$$

Eq. (22) gives the two EPs

$$K_\pm = \frac{2\varepsilon_o \pm \varepsilon_r}{4\varepsilon_o \pm \varepsilon_r}\sqrt{\varepsilon_o}\Omega, \tag{23}$$

at which the two bands coalesce into a single defective quasienergy $Q = \Omega / 2$. The existence of EPs is a result of non-Hermitian off-diagonal terms in Eq. (21). From Eq. (23), we find to the first order in $\varepsilon_r / \varepsilon_o$, the size of the k-gap as

$$\Delta K = K_+ - K_- \approx \frac{\varepsilon_r}{4\varepsilon_o}\sqrt{\varepsilon_o}\Omega. \tag{24}$$

At the EPs, the defective eigenstates obtained from Eq. (21) have the following forms:

$$\left|\tilde{\psi}_{-}^{R}\right\rangle = \begin{pmatrix} c_2^1 \\ c_1^0 \end{pmatrix} \propto \begin{pmatrix} -ie^{i\phi} \\ 1 \end{pmatrix}, \qquad \left|\tilde{\psi}_{+}^{R}\right\rangle = \begin{pmatrix} c_2^1 \\ c_1^0 \end{pmatrix} \propto \begin{pmatrix} ie^{i\phi} \\ 1 \end{pmatrix}. \tag{25}$$

The chirality of EP can be defined analogous to the polarization of electromagnetic waves [48]. Eq. (25) shows clearly that the two EPs possess opposite chirality, which is expected because they originate from the same diabolic point. The chirality is determined by the initial phase $\phi$ or the corresponding initial time $t_0 = \phi / \Omega$. So the chirality varies periodically as $t_0$ changes and the stationary chirality can be observed by a discrete measurement with the time step equals to $2\pi / \Omega$. Substituting the expression of $\tilde{\psi}_1^R$ into Eq. (11), we obtain the transformed Floquet mode at the EP $K = K_-$ as

$$\vec{\Phi}' = c_2^1 \vec{\varphi}_2 e^{i\Omega t} + c_1^0 \vec{\varphi}_1 \propto -ie^{i\phi} \begin{pmatrix} 0 \\ e^{-iK_- x} e^{i\Omega t} \end{pmatrix} + \begin{pmatrix} e^{iK_- x} \\ 0 \end{pmatrix}. \tag{26}$$

In order to find the explicit eigenfunction of the magnetic field at the EPs, different from Eq. (25), we have chosen the first component of the wavefunction in Eq. (26) as the positive band, whereas the second component is chosen as the negative band. In the limit of $|\varepsilon_r / \varepsilon_o| \ll 1$, and $K_- \approx K_+ \approx \sqrt{\varepsilon_o}\Omega / 2$, the time-dependent magnetic field and its derivative can be obtained from the inverse of the transformation defined in Eq. (9), i.e.,

$$\begin{pmatrix} H_z \\ \dot{H}_z \end{pmatrix} = e^{-i\Omega t/2} \vec{\Phi} = e^{-i\Omega t/2} R^{-1} \cdot \vec{\Phi}' \propto \frac{1}{2} \begin{pmatrix} \frac{2}{\Omega} (e^{iK_- x} e^{-i\Omega t/2} - ie^{i\phi} e^{-iK_- x} e^{i\Omega t/2}) \\ -ie^{iK_- x} e^{-i\Omega t/2} + e^{i\phi} e^{-iK_- x} e^{i\Omega t/2} \end{pmatrix}. \tag{27}$$

Similarly, we can obtain the magnetic field and its derivative with respect to time at the EP $K = K_+$ as

$$\begin{pmatrix} H_z \\ \dot{H}_z \end{pmatrix} \propto \frac{1}{2} \begin{pmatrix} \frac{2}{\Omega} (e^{iK_+ x} e^{-i\Omega t/2} + ie^{-i\phi} e^{-iK_+ x} e^{i\Omega t/2}) \\ -ie^{iK_+ x} e^{-i\Omega t/2} - e^{-i\phi} e^{-iK_+ x} e^{i\Omega t/2} \end{pmatrix}. \tag{28}$$

Eqs. (27) and (28) show that the propagating field possesses two terms; the first term represents the ordinary wave propagation due to the zeroth order positive band, whereas the second term comes from the negative band of the first order. And the chirality of the EPs determines the phase difference between these two terms.

To identify the singularity of the EPs, we studied the phase rigidity $r_\pm$ at the EPs, which is defined as $r_\pm = \langle \tilde{\psi}_\pm^L | \tilde{\psi}_\pm^R \rangle$ [49], where $\langle \tilde{\psi}_\pm^L |$ and $| \tilde{\psi}_\pm^R \rangle$ are the left and right eigenstates of the non-Hermitian Hamiltonian Eq. (21), respectively, and the bracket denotes the inner product. Since the Hamiltonian $\tilde{\mathrm{H}}_{F,red}$ is non-Hermitian, the right eigenstates $| \tilde{\psi}_\pm^R \rangle$ are different from the left eigenstates $\langle \tilde{\psi}_\pm^L |$. Both together form a biorthogonal basis [48]. At the EPs, the left eigenstates can be calculated according to

$$\langle \tilde{\psi}_\pm^L | \tilde{\mathrm{H}}_{F,red} = \frac{\Omega}{2} \langle \tilde{\psi}_\pm^L |, \tag{29}$$

and one arrives at

$$\langle \tilde{\psi}_-^L | \propto \begin{pmatrix} -ie^{-i\phi} & 1 \end{pmatrix}, \qquad \langle \tilde{\psi}_+^L | \propto \begin{pmatrix} ie^{-i\phi} & 1 \end{pmatrix}. \tag{30}$$

Phase rigidity vanishes at the EP according to a power-law behavior. Combining Eq. (25) and Eq. (30), we can easily find that the phase rigidities at the two EPs are indeed zero.

Similarly, two opposite bands with orders differing by any integer $\pm n$ can be coupled by $\tilde{\mathrm{H}}_{\pm n}$ and create infinite number of pairs of EPs around $K = n\sqrt{\varepsilon_o}\Omega/2$ with quasienergies at $Q = m\Omega, m \in \mathbb{Z}$. For example, to find the second-order EPs at $Q = 0$, we construct a 14×14 reduced Floquet Hamiltonian from Eq. (20) centered at $\tilde{\mathrm{H}}_0$, from which, we find another pair of EPs located near $K = \sqrt{\varepsilon_o}\Omega$, as shown in the inset of Fig. 1b. However, the size of the second-order k-gap is much smaller than that of the first order. Similar pairs appear periodically at $Q = m\Omega$ for all non-zero integers *m*. There are infinite number of pairs of EPs with decreasing size of the k-gap as K increases. For a 3D homogeneous Floquet medium, these EPs span infinite number of pairs of spherical surfaces all centered at $K=0$ with radii around $K=n\sqrt{\varepsilon_o}\Omega/2$ with $n$ being either even or odd.

In order to verify the existence of a broken phase between a pair of EPs, we use the FDTD method to numerically simulate the electromagnetic (EM) wave propagation inside a slab of dynamic medium embedded in air with thickness $L$ and a time-dependent relative permittivity $\varepsilon(t)=\varepsilon_o+\varepsilon_r\sin(\Omega t+\phi)$. The medium is located between $0 < x < L$ and we excite the magnetic field by using a pulsed line source of the form $\exp(-0.5(t-t_0)^2/\tau^2)\delta(x-x_0)$ with $x_0 < 0$, where $\tau$ is the duration time. As shown in Fig. 2, we discretized the medium to $N+1$ spatial steps and $M+1$ temporal steps with the step sizes $\Delta x$ and $\Delta t$, respectively. Then the time-dependent electromagnetic fields can be calculated using the following discretized Maxwell's equations,

$$E_y\Big|_j^{i+1}=\frac{\varepsilon\big|_j^i}{\varepsilon\big|_j^{i+1}}E_y\Big|_j^i-\frac{\Delta t}{\varepsilon\big|_j^{i+1}\varepsilon_0}\frac{H_z\Big|_{j+1/2}^{i+1/2}-H_z\Big|_{j-1/2}^{i+1/2}}{\Delta x},$$

$$H_z\Big|_{j+1/2}^{i+1/2}=H_z\Big|_{j+1/2}^{i-1/2}-\frac{\Delta t}{\mu_0}\frac{E_y\Big|_{j+1}^{i}-E_y\Big|_j^i}{\Delta x}.$$

The magnetic field inside the slab is written as

$$H_z(t,x)=\int_{-\infty}^{\infty}e^{-iQ(K)t}\sum_{n=-\infty}^{\infty}f_K^n e^{iKx}e^{in\Omega t}dK, \tag{31}$$

where $f_K^n$ is the amplitude corresponding to wave vector $K$ and the quasienergies in a band of order $n$. Note that here the quasienergy are chosen in the first Brillouin zone $-\Omega/2\le Q\le\Omega/2$. As we discussed previously, when $\varepsilon_r$ is real, there are infinite number of pairs of EPs as $K$ increases and between each pair of EPs is a broken phase with the quasienergies in complex conjugate pairs. As the time grows, according to Eq. (6), the field components inside the slab corresponding to the Floquet states whose quasienergies possess negative imaginary parts will decay while the field components corresponding to the Floquet states whose quasienergies possess positive imaginary parts will grow. At large times, the fields inside the slab will be dominated by the field components whose wave vectors are inside the broken phase region. As a result, this kind of dynamic medium can induce a field amplification [35, 36] if the medium is excited at the proper K.

To describe the amplification behavior, we defined

$$g_0(K) = \frac{1}{(t_2 - t_1)L} \int_0^L e^{-iKx} dx \int_{t_1}^{t_2} H_z e^{i\Omega t/2} dt, \tag{32}$$

where $t_1$, $t_2$ are the starting and ending time for the measurement. According to Eq. (31), $g_0(K)$ equals to $f_K^0$ in the condition that $Q = \Omega/2$. In Fig. 3 (a) and (b), we plot the imaginary parts of the quasienergies as well as $|g_0(K)|$ as functions of the wave vector $K$. We can see clearly that for the wave vectors in the broken phase region where the imaginary parts of the quasienergies are non-zero, $g_0(K)$ becomes extremely large when the integration time is long enough. Since the amplitude of the incident field is less than 1, the field inside the slab is greatly amplified. And comparing Figs. 3(a) and 3(b), we see that when the range of the broken phase decreases by decreasing $\varepsilon_r$, the region of wave vectors for the field amplification also becomes narrower. As the positive imaginary part of the quasienergy represents the growing rate of the Floquet states, a larger imaginary part corresponds to a stronger the amplification. To show the periodicity of the Floquet bands, we also defined

$$g_1(K) = \frac{1}{(t_2 - t_1)L} \int_0^L e^{-iKx} dx \int_{t_1}^{t_2} H_z e^{i3\Omega t/2} dt, \tag{33}$$

and plotted $g_1(K)$ as function of $K$ in Fig. 3(c). $g_1(K)$ denotes the amplitude of Floquet state corresponding to $Q = 3\Omega/2$. We can see that $g_1(K)$ has the identical shape as that in Fig. 2(b) and is also largely amplified in the broken phase region.

If we increase both the starting time $t_1$ and ending time $t_2$ by $\Delta t$, according to Eqs. (31) and (32), $g_0(K)$ must be amplified by $\exp[\mathrm{Im}(Q)\Delta t]$. We can therefore numerically determine the imaginary part of the quasienergy according to the amplifications for different $\Delta t$. In Fig.4(a), we showed the logarithm of the maximum $|g_0(K)|$ (corresponding to the center of the *k*-gap) as function of $\Delta t$ for different modulation permittivity $\varepsilon_r$. It is clearly seen that for each $\varepsilon_r$, $\ln|g_0(K)|$ is linear in $\Delta t$, and its slope $b$ determines the maximum imaginary quasienergy according to $\text{Im}(Q_m)/\Omega = b/(2\pi)$. The maximum imaginary quasienergy $\text{Im}(Q_m)$ as function of the modulation permittivity $\varepsilon_r$ is shown in Fig. 4(b), in which $\text{Im}(Q_m)$ is calculated using both the Floquet Hamiltonian (in solid line) and linear fitting of $\ln|g_0(K)| \sim \Delta t$ (in red circles). The excellent agreement between the two results is clearly seen. The linear relation between $\text{Im}(Q_m)$ and $\varepsilon_r$ can also be seen from Eq. (22), i.e., $\text{Im}(Q_m) = \varepsilon_r \Omega/(8\varepsilon_o)$ at the k-gap center $K = \sqrt{\varepsilon_o}\Omega/2$. The linear relation between $\text{Im}(Q_m)$ and $\varepsilon_r$ found above is analogous to the Bragg scattering in k-space when a gap is open at the Brillouin zone boundary due to a periodic potential of strength $\Delta V$, the imaginary part of k at mid-gap is also proportional to $\Delta V$.

## IV. Band gaps

When $\varepsilon_r$ is purely imaginary, the reduced Hamiltonian $\tilde{\text{H}}_{F,red}$ in Eq. (21) becomes Hermitian. In this case, it is expected that the interaction between two opposite bands in the vicinity of the diabolic points at $K = \sqrt{\varepsilon_o}\Omega/2$ will become repulsive and form a quasi-energy gap. Indeed, when $\varepsilon_r^2$ in Eq. (22) is replaced by $-|\varepsilon_r^2|$, we find two separated bands with the lower band edge at $Q = Q_-$ and the upper band edge at $Q = Q_+$ as shown in Fig. 1(c). The gap size $\Delta Q = Q_+ - Q_-$ can be calculated from Eq. (22). At $K = \sqrt{\varepsilon_o}\Omega/2$, we find

$$\Delta Q = \sqrt{\left(\frac{2K}{\sqrt{\varepsilon_o}} - \Omega\right)^2 + \frac{|\varepsilon_r^2|}{4\varepsilon_o^2}\left(\frac{K}{\sqrt{\varepsilon_o}} - \Omega\right)^2}\Bigg|_{K=\sqrt{\varepsilon_o}\Omega/2} = \frac{|\varepsilon_r|\Omega}{4\varepsilon_o}. \tag{34}$$

If we defined $\varepsilon_r = i\tilde{\varepsilon}_r$, with $\tilde{\varepsilon}_r$ being a real number, it can be seen from Eq. (34) that the gap will close and reopen when $\tilde{\varepsilon}_r$ changes from a positive value to a negative one. When $\tilde{\varepsilon}_r > 0$, the eigenvectors of Eq. (22) corresponding to $Q = Q_-$ and $Q = Q_+$ are, respectively

$$\tilde{\psi}_-^R = \begin{pmatrix} c_2^1 \\ c_1^0 \end{pmatrix} \propto \begin{pmatrix} e^{i\phi} \\ 1 \end{pmatrix}, \qquad \tilde{\psi}_+^R = \begin{pmatrix} c_2^1 \\ c_1^0 \end{pmatrix} \propto \begin{pmatrix} -e^{i\phi} \\ 1 \end{pmatrix}. \tag{35}$$

Substituting Eq. (35) into Eq. (11) and combining Eq. (6), we obtain the following magnetic fields and their time derivatives at $Q = Q_-$ and $Q = Q_+$, respectively,

$$\begin{pmatrix} H_z \\ \dot{H}_z \end{pmatrix} = e^{-iQ_- t}\hat{R}^{-1}\cdot\vec{\Phi}' \propto e^{-iQ_- t}\begin{pmatrix} \frac{2}{\Omega}(e^{iKx}+e^{i\phi}e^{-iKx}e^{i\Omega t}) \\ -ie^{iKx}+ie^{i\phi}e^{-iKx}e^{i\Omega t} \end{pmatrix} \approx e^{i\phi/2}\begin{pmatrix} \frac{4}{\Omega}\cos(\frac{\sqrt{\varepsilon_o}\Omega x-\Omega t-\phi}{2}) \\ 2\sin(\frac{\sqrt{\varepsilon_o}\Omega x-\Omega t-\phi}{2}) \end{pmatrix}, \quad (36)$$

and

$$\begin{pmatrix} H_z \\ \dot{H}_z \end{pmatrix} = e^{-iQ_+ t}\hat{R}^{-1}\cdot\vec{\Phi}' \propto e^{-iQ_+ t}\begin{pmatrix} \frac{2}{\Omega}(e^{iKx}-e^{i\phi}e^{-iKx}e^{i\Omega t}) \\ -ie^{iKx}-ie^{i\phi}e^{-iKx}e^{i\Omega t} \end{pmatrix} \approx ie^{i\phi/2}\begin{pmatrix} \frac{4}{\Omega}\sin(\frac{\sqrt{\varepsilon_o}\Omega x-\Omega t-\phi}{2}) \\ -2\cos(\frac{\sqrt{\varepsilon_o}\Omega x-\Omega t-\phi}{2}) \end{pmatrix}, \quad (37)$$

where we have assumed $\varepsilon_r / \varepsilon_o << 1$ so that $\Delta Q \ll 1$. Eqs. (36) and (37) show that the magnetic fields for the Floquet states at the two band edges are standing waves in the stroboscopic sense, similar to the Bloch states at the band edges for the one-dimensional ordinary photonic crystal. It is interesting to point out that the two solutions shown in Eqs. (36) and (37) differ by a time-shift $\pi / \Omega$, which is a half period of the modulation frequency $\Omega$. Again, this is analogous to the spatial phase shift $\Lambda / 2$ of two orthogonal standing waves at the band edges of the Bragg scattering gap, which is also half of the spatial period $\Lambda$.

We have also numerically simulated the time-dependent fields to obtain the band structures close to the gap. Similar to the simulation of broken phase discussed earlier for real $\varepsilon_r$, we use again a pulsed line source of the magnetic field of the form $\exp(-0.5(t-t_0)^2/\tau^2)$ located at $x = x_0$ to the left to a dynamic slab of thickness $L$ with a relative permittivity of the form $\varepsilon(t) = 5 + 1.5i\sin\Omega t$. From Eq. (31), the magnetic field inside the slab for a specific wave vector $K$ and frequency $\omega$ can be obtained through the Fourier Transform

$$h(K,\omega) = \frac{1}{(t_2-t_1)L}\int_{t_1}^{t_2} dt e^{i\omega t}\int_0^L H_z e^{-iKx}dx. \quad (38)$$

According to Eq. (31), $h(K,\omega)$ is not zero only when $\omega$ equals to $Q+n\Omega$, where $Q$ is the quasienergy corresponding to $K$ and $n$ is an arbitrary integer number.

In Fig. 5(a), we plot the magnitude of $h(K,\omega)$ for the fixed wave vector $K=1.1\Omega$ as function of the frequency $\omega$ in the range of $0<\omega<\Omega$. Two peaks are found at $\omega\approx 0.46\Omega$ and $\omega\approx 0.54\Omega$. In order to obtain the band structures, we repeated the calculations for other values of $K$ in the region of $0.9\Omega\le K\le 1.4\Omega$. The results are shown by circles in Fig. 5(b). The solid lines are the results obtained from the Floquet Hamiltonian Eq. (20). The excellent agreement between the two is clearly seen. Similar to the case of real $\varepsilon_r$, we have also studied the second order effect by considering the coupling of two opposite bands with orders differing by $\pm 2$ due to $\tilde{\mathrm{H}}_{\pm 2}$. Although $\varepsilon_r$ is pure imaginary, its square becomes a real number as can be seen from Eq. (19). Thus, pairs of EPs are expected at $K=\sqrt{\varepsilon_o}\Omega$ and quasienergies $Q=m\Omega,\ m\in\mathbb{Z}$. For the case of $Q=0$, we construct a $14\times 14$ reduced Floquet Hamiltonian from Eq. (20) centered at $\tilde{\mathrm{H}}_0$, from which, we indeed find another pair of EPs located near $K=\sqrt{\varepsilon_o}\Omega$ as shown in the inset of Fig. 1c. The similar pairs appear periodically at $Q=m\Omega$ for all non-zero integers *m*.

We have also studied the case that $\varepsilon_r$ is a complex number. According to Eq. (22), when $\varepsilon_r$ is a complex number, there will be no EPs as the formula in the square root can no longer be zero. In this case, the singularity will be smoothed [50, 51]. The quasienergies are now complex numbers with the imaginary parts can be both positive and negative. And there are gaps for the real part of the quasienergies. These can be verified by calculating the Floquet bands using the Floquet Hamiltonian as shown in Fig 6.

**V. Space-time modulated permittivity**

In this section, we will study the more complicated case that the permittivity is modulated simultaneously in space and time, i.e., the permittivity is expressed as

$$\varepsilon(x,t)=\varepsilon_o+\varepsilon_r(\Omega t-\beta x+\phi), \tag{39}$$

where $\beta$ and $\Omega$ are the spatial and temporal modulation frequencies, respectively. The modulation speed is defined as $c_m = \Omega / \beta$. The scattering of light by a space-time modulated medium is widely studied in acousto-optics [52-54]. By illuminating an acoustic plane wave on the $TeO_2$ crystal, the modulation permittivity of the crystal can be expressed as Eq. (39) due to the elasto-optic effect. The diffraction of light by the space-time modulated medium can be calculated by directly solving the Maxwell's equations, and the diffraction efficiency of diffracted orders has been studied by Laude using the coupled-wave equations method [52]. The scattering of acoustic wave by a space-time modulated medium has also been studied recently [55]. And it is shown that *k*-gap forms when the modulation speed is faster than the speed of wave inside the medium. In the following, we will show that this conclusion is also valid for the electromagnetic system, and a pair of EPs forms at the edges of the *k*-gap.

We first numerically calculate the band dispersion of the space-time modulated medium using the plane wave expansion method. Based on the Floquet-Bloch theorem, the electromagnetic fields inside the space-time modulated medium can be written as

$$E_y = e^{iKx-iQt} \sum_{n=-\infty}^{\infty} E_{yn} e^{in(\Omega t-\beta x)}, \quad H_z = e^{iKx-iQt} \sum_{n=-\infty}^{\infty} H_{zn} e^{in(\Omega t-\beta x)}. \tag{40}$$

And the space-time modulated permittivity can be expressed as

$$\varepsilon(x,t) = \varepsilon_o + \varepsilon_1 e^{i(\Omega t-\beta x)} + \varepsilon_{-1} e^{-i(\Omega t-\beta x)}, \tag{41}$$

where $\varepsilon_{\pm 1} = e^{\pm i\phi} \varepsilon_r / 2$. Substituting Eqs. (40) and (41) into the Maxwell's equations, one arrives at

$$\begin{aligned}
& i\sum_{n=-\infty}^{\infty} (K-n\beta) E_{yn} e^{iKx-iQt} e^{in(\Omega t-\beta x)} = i\sum_{n=-\infty}^{\infty} (Q-n\Omega) H_{zn} e^{iKx-iQt} e^{in(\Omega t-\beta x)}, \\
& i\sum_{n=-\infty}^{\infty} (K-n\beta) H_{zn} e^{iKx-iQt} e^{in(\Omega t-\beta x)} = \sum_{m=-\infty}^{\infty} \varepsilon_m e^{im(\Omega t-\beta x)} \sum_{n=-\infty}^{\infty} -i(Q-n\Omega) E_{yn} e^{iKx-iQt} e^{in(\Omega t-\beta x)} \\
& +\sum_{m=-\infty}^{\infty} im\Omega \varepsilon_m e^{im(\Omega t-\beta x)} \sum_{n=-\infty}^{\infty} E_{yn} e^{ikx-iQt} e^{in(\Omega t-\beta x)}.
\end{aligned} \tag{42}$$

Eq. (42) leads to the following eigenvalue problem with the wave vector *K* as the eigenvalues

$$K\begin{pmatrix} \vdots \\ E_{y0} \\ H_{z0} \\ \vdots \end{pmatrix} = \hat{\mathrm{H}} \cdot \begin{pmatrix} \vdots \\ E_{y0} \\ H_{z0} \\ \vdots \end{pmatrix}, \tag{43}$$

where $\hat{\mathrm{H}}$ is defined as

$$\begin{pmatrix} \ddots & \ddots & & & & & \\ \ddots & \hat{\mathrm{H}}_{-2} & \varepsilon_{-1}\hat{\Delta}_{-2} & & & & \\ & \varepsilon_1\hat{\Delta}_{-1} & \hat{\mathrm{H}}_{-1} & \varepsilon_{-1}\hat{\Delta}_{-1} & & & \\ & & \varepsilon_1\hat{\Delta}_0 & \hat{\mathrm{H}}_0 & \varepsilon_{-1}\hat{\Delta}_0 & & \\ & & & \varepsilon_1\hat{\Delta}_1 & \hat{\mathrm{H}}_1 & \varepsilon_{-1}\hat{\Delta}_1 & \\ & & & & \varepsilon_1\hat{\Delta}_2 & \hat{\mathrm{H}}_2 & \ddots \\ & & & & & \ddots & \ddots \end{pmatrix}, \tag{44}$$

with

$$\hat{\mathrm{H}}_n = \begin{pmatrix} n\beta & \omega - n\Omega \\ (\omega - n\Omega)\varepsilon_o & n\beta \end{pmatrix}, \qquad \hat{\Delta}_n = \begin{pmatrix} 0 & 0 \\ \omega - n\Omega & 0 \end{pmatrix}. \tag{45}$$

The Floquet band dispersion is obtained by solving the eigenvalue equation (43) for every given quasienergy $Q$.

In Fig. 7, we showed the Floquet bands in the region $0 \le Q \le \Omega, -\beta/2 \le K \le \beta/2$ for the space-time modulated medium with different modulation speeds, which are calculated using Eq. (43) with $\hat{\mathrm{H}}$ is truncated at $n$ = 22. From Eq. (43), in the limit of $|\varepsilon_r/\varepsilon_o| \to 0$ where the coupling matrices $\hat{\Delta}_n$ vanish, we can see that the Floquet bands are obtained by copy the two linear bands of $\hat{\mathrm{H}}_0$ and shifting them up and down in the $(K,Q)$ by integers multiplying $(\beta,\Omega)$. Similarly, we call the bands formed by shifting the bands of $\hat{\mathrm{H}}_0$ by $(n\beta, n\Omega)$ as band $n$. There are intersections between bands of different orders. When $\varepsilon_r \neq 0$, $k$-gaps or quasienergy gaps will form around the intersections due to the couplings between the bands of different orders. When the modulation speed $(c_m = \Omega/\beta)$ is slower than the wave speed inside the medium which is $c/\sqrt{\varepsilon_o} \approx 0.45$, quasienergy gaps form around the intersections, as shown in Fig 7(a). However, when the modulation speed is faster than the wave speed, *k*-gaps are created as can be seen in Fig 7(b). To investigate the EP phenomena associated with the k-gaps, we use again the Floquet matrix method to calculate the Floquet bands for the case of $c_m > c/\sqrt{\varepsilon_o}$, and show that the region inside the *k*-gap is the broken phase.

As the permittivity depends on $(x,t)$ through a unique combination, the medium is more conveniently described by a new viable *u* introduced below

$$(x,u) = (x, t - x/c_m), \qquad (\partial_x, \partial_t) = (\partial_x - 1/c_m \partial_u, \partial_u). \tag{46}$$

According to the Maxwell's equations, the wave equation for the electric field inside the space-time modulated medium is written as

$$-\frac{\partial^2 E}{\partial x^2} = -\frac{\partial}{\partial t^2}(\varepsilon(t)E). \tag{47}$$

Using the new variables $(x,u)$ and noting that $\varepsilon(x,t) = \varepsilon(u) = \varepsilon_o + \varepsilon_r \cos(\Omega u + \phi)$, we rewrite Eq. (47) as

$$-(\frac{\partial^2}{\partial x^2} - \frac{2}{c_m}\frac{\partial}{\partial x}\frac{\partial}{\partial u} + \frac{1}{c_m^2}\frac{\partial^2}{\partial u^2})E_y = -\frac{\partial^2}{\partial u^2}(\varepsilon(u)E_y) = -\varepsilon(u)\frac{\partial^2 E_y}{\partial u^2} - 2\frac{\partial \varepsilon(u)}{\partial u}\frac{\partial E_y}{\partial u} - \frac{\partial^2 \varepsilon(u)}{\partial u^2}E_y. \tag{48}$$

If we define $\tilde{K} = K - Q/c_m$, according to Eq. (40), the electric field can be written as

$$E_y = e^{iKx - iQt}\sum_{n=-\infty}^{\infty} q_n e^{in(\Omega t - \beta x)} = e^{iKx - iQ(u + x/c_m)}\sum_{n=-\infty}^{\infty} q_n e^{in(\Omega t - \beta x)} = e^{i\tilde{K}x - iQt}\sum_{n=-\infty}^{\infty} q_n e^{in\Omega u}, \tag{49}$$

where $q_n$ is the expansion coefficients of order *n*. Substituting Eq. (49) into Eq. (48), one arrives at

$$-(-\tilde{K}^2 - \frac{2}{c_m}i\tilde{K}\frac{\partial}{\partial u} + \frac{1}{c_m^2}\frac{\partial^2}{\partial u^2})E_y = -\frac{\partial^2}{\partial u^2}(\varepsilon(u)E_y) = -\varepsilon(u)\frac{\partial^2 E_y}{\partial u^2} - 2\frac{\partial \varepsilon(u)}{\partial u}\frac{\partial E_y}{\partial u} - \frac{\partial^2 \varepsilon(u)}{\partial u^2}E_y. \tag{50}$$

Employing the Liouvillian formulation, we can rewrite Eq. (50) as a two-component time-dependent Schrodinger-like equation for the electric field and its derivative with respect to $u$ as

$$i\frac{\partial}{\partial u}\begin{pmatrix} E \\ \dot{E} \end{pmatrix} = \begin{pmatrix} 0 & i \\ -iB' & -iA' \end{pmatrix}\begin{pmatrix} E \\ \dot{E} \end{pmatrix} = \hat{\mathrm{H}}'_{eff}\begin{pmatrix} E \\ \dot{E} \end{pmatrix}, \tag{51}$$

where $\dot{E} = \partial E / \partial u$ and

$$A' = \frac{2i\tilde{K}/c_m + 2\partial\varepsilon/\partial u}{\varepsilon(u) - 1/c_m^2}, \qquad B' = \frac{\tilde{K}^2 + \partial^2\varepsilon/\partial u^2}{\varepsilon(u) - 1/c_m^2}. \tag{52}$$

Similarly, we apply the following similarity transformation to the effective Hamiltonian $\hat{\mathrm{H}}'_{eff}$ so that its time-independent part $\tilde{\mathrm{H}}'_0$ becomes Hermittian:

$$\tilde{\mathrm{H}}'(u) = \hat{R}'\cdot\hat{\mathrm{H}}'_{eff}(u)\cdot\hat{R}'^{-1} = \begin{pmatrix} \dfrac{B' - iA'\sqrt{C'} + C'}{2\sqrt{C'}} & \dfrac{B' + iA'\sqrt{C'} - C'}{2\sqrt{C'}} \\ \dfrac{-B' + iA'\sqrt{C'} + C'}{2\sqrt{C'}} & \dfrac{-B' - iA'\sqrt{C'} - C'}{2\sqrt{C'}} \end{pmatrix}, \tag{53}$$

where

$$\hat{R}' = \begin{pmatrix} \sqrt{C'} & i \\ \sqrt{C'} & -i \end{pmatrix}, \qquad C' = \frac{\tilde{K}^2}{\varepsilon_o - 1/c_m^2}. \tag{54}$$

When $\varepsilon_r = 0$, the time-independent Hamiltonian becomes

$$\tilde{\mathrm{H}}'_0 = \begin{pmatrix} \dfrac{c_m\tilde{K}}{c_m^2\varepsilon_o - 1} + \sqrt{\dfrac{c_m^2\tilde{K}^2}{c_m^2\varepsilon_o - 1}} & \dfrac{c_m\tilde{K}}{1 - c_m^2\varepsilon_o} \\ \dfrac{c_m\tilde{K}}{1 - c_m^2\varepsilon_o} & \dfrac{c_m\tilde{K}}{c_m^2\varepsilon_o - 1} - \sqrt{\dfrac{c_m^2\tilde{K}^2}{c_m^2\varepsilon_o - 1}} \end{pmatrix}. \tag{55}$$

We can see that $\tilde{\mathrm{H}}'_0$ is Hermittian when $c_m^2 > 1/\varepsilon_o = c^2/\varepsilon_o$. From Eq. (55), we obtain the band dispersion of $\tilde{\mathrm{H}}'_0$ as

$$Q = \frac{c_m \tilde{K} \pm c_m^2 \sqrt{\varepsilon_o} \tilde{K}}{c_m^2 \varepsilon_o - 1}. \quad (56)$$

Since $\tilde{K} = K - Q / c_m$, Eq. (56) can be rewritten in a simpler form:

$$Q = \pm K / \sqrt{\varepsilon_o}, \quad (57)$$

which is the same as the band dispersion of $\tilde{\mathrm{H}}_0$. In the limit of small $|\varepsilon_r / \varepsilon_o|$, we can expand the new time-dependent effective Hamiltonian $\tilde{\mathrm{H}}'(u)$ in Taylor series. To the first order of $|\varepsilon_r / \varepsilon_o|$, $\tilde{\mathrm{H}}'(u)$ can be expressed as

$$\tilde{\mathrm{H}}'(u) = \tilde{\mathrm{H}}'_0 + \tilde{\mathrm{H}}'_1 e^{i\Omega u} + \tilde{\mathrm{H}}'_{-1} e^{-i\Omega u}, \quad (58)$$

where

$$\tilde{\mathrm{H}}'_{\pm 1} = \begin{pmatrix} H_{11}^{\pm} & H_{12}^{\pm} \\ H_{21}^{\pm} & H_{22}^{\pm} \end{pmatrix}, \quad (59)$$

with

$$\begin{aligned}
H_{11}^{\pm} &= ie^{\pm i\phi} \varepsilon_r \frac{\pm \tilde{K}^2 \pm 2\tilde{K}^2 \sqrt{\frac{1}{c_m^2 \varepsilon_o - 1}} + (\frac{1}{c_m^2} - \varepsilon_o)(2 c_m \tilde{K} \sqrt{\frac{1}{c_m^2 \varepsilon_o - 1}} \mp \Omega)\Omega}{4(\frac{1}{c_m^2} - \varepsilon_o)^2 c_m \tilde{K} \sqrt{\frac{1}{c_m^2 \varepsilon_o - 1}}}, \\
H_{12}^{\pm} &= ie^{\pm i\phi} \varepsilon_r \frac{\pm \tilde{K}^2 \mp 2\tilde{K}^2 \sqrt{\frac{1}{c_m^2 \varepsilon_o - 1}} - (\frac{1}{c_m^2} - \varepsilon_o)(2 c_m \tilde{K} \sqrt{\frac{1}{c_m^2 \varepsilon_o - 1}} \pm \Omega)\Omega}{4(\frac{1}{c_m^2} - \varepsilon_o)^2 c_m \tilde{K} \sqrt{\frac{1}{c_m^2 \varepsilon_o - 1}}}, \\
H_{21}^{\pm} &= -H_{11}^{\pm}, \qquad H_{22}^{\pm} = -H_{12}^{\pm}.
\end{aligned} \quad (60)$$

Following the same procedure as we did in section II, we can then obtain the Floquet Hamiltonian $\tilde{\mathrm{H}}'_F$ as

$$\tilde{\mathrm{H}}'_F = \begin{pmatrix} \ddots & \ddots & 0 & 0 & 0 \\ \ddots & \tilde{\mathrm{H}}'_0 + \Omega \overleftrightarrow{\mathrm{I}}_2 & \tilde{\mathrm{H}}'_1 & 0 & 0 \\ 0 & \tilde{\mathrm{H}}'_{-1} & \tilde{\mathrm{H}}'_0 & \tilde{\mathrm{H}}'_1 & 0 \\ 0 & 0 & \tilde{\mathrm{H}}'_{-1} & \tilde{\mathrm{H}}'_0 - \Omega \overleftrightarrow{\mathrm{I}}_2 & \ddots \\ 0 & 0 & 0 & \ddots & \ddots \end{pmatrix}. \quad (61)$$

When $\varepsilon_r$ is larger, we should also keep the higher order terms $\tilde{\mathrm{H}}'_{\pm n}$ with $n > 1$ and consider the corresponding couplings in the Floquet Hamiltonian matrix.

In Fig. 8(a), we showed the real part of the Floquet bands calculated using the Floquet matrix method (circles) and those calculated using the plane wave expansion method (lines). We can see that the Floquet bands calculated using these two methods agree excellently. What's more, using the Floquet matrix method, the imaginary part of the quasienergy as a function of the wavenumber is also calculated and shown in Fig. 8(b). It is clearly seen that the quasienergies inside the *k*-gap are complex conjugate pairs, indicating that the region inside the *k*-gap is broken phase. Note that to calculate the Floquet bands in Fig. 8, we expanded $\tilde{\mathrm{H}}'(u)$ and kept up to the third order terms (the coupling matrices $\tilde{\mathrm{H}}'_{\pm 2}, \tilde{\mathrm{H}}'_{\pm 3}$ are considered in the Floquet matrix).

## VI. Conclusions

In summary, we have investigated the formation of EPs in a time-Floquet photonic system in which the permittivity of the dynamic medium is periodic in time. Using the method of Floquet matrix, we obtained the photonic Floquet band dispersions analytically and the expressions of the Floquet states in the limit of $|\varepsilon_r / \varepsilon_o| \to 0$. In photonics, the permittivity is the classical counterpart of the potential in a quantum system. However, the time modulated permittivity $\varepsilon_r$ performs very differently from the time modulated potential in quantum periodically driven systems. When $\varepsilon_r$ is a real number, the time modulation induces infinitely many pairs of exceptional points, each pair spanning two concentric spherical surfaces in the three dimensional wave vector space. Those surfaces of EPs also repeat themselves in the quasienergy space with the modulation frequency as the period. Since the Floquet states in the broken phase region will decay and grow as the time increases, the time modulation of permittivity can induce the field amplification and damping in the absence of loss and gain materials. When $\varepsilon_r$ is purely imaginary, there will be quasienergy gaps with the gap sizes proportional to $|\varepsilon_r / \varepsilon_o|$. For a complex $\varepsilon_r$, there is no EP but since the quasienegies form complex conjugate pairs, field amplification and damping still exist. We have also extended our theory to the more complicated case where the permittivity depends on both space and time through a unique combination. It is shown that EPs still exist for a real $\varepsilon_r$ when the modulation speed is faster than the speed of wave inside the medium.

**Acknowledgements:** We thank Drs. Kun Ding and Ruoyang Zhang for stimulating discussions and useful suggestions. This work is supported by Hong Kong Research Grants Council through grant No. AoE/P-02/12 and N_HKUST608/17.

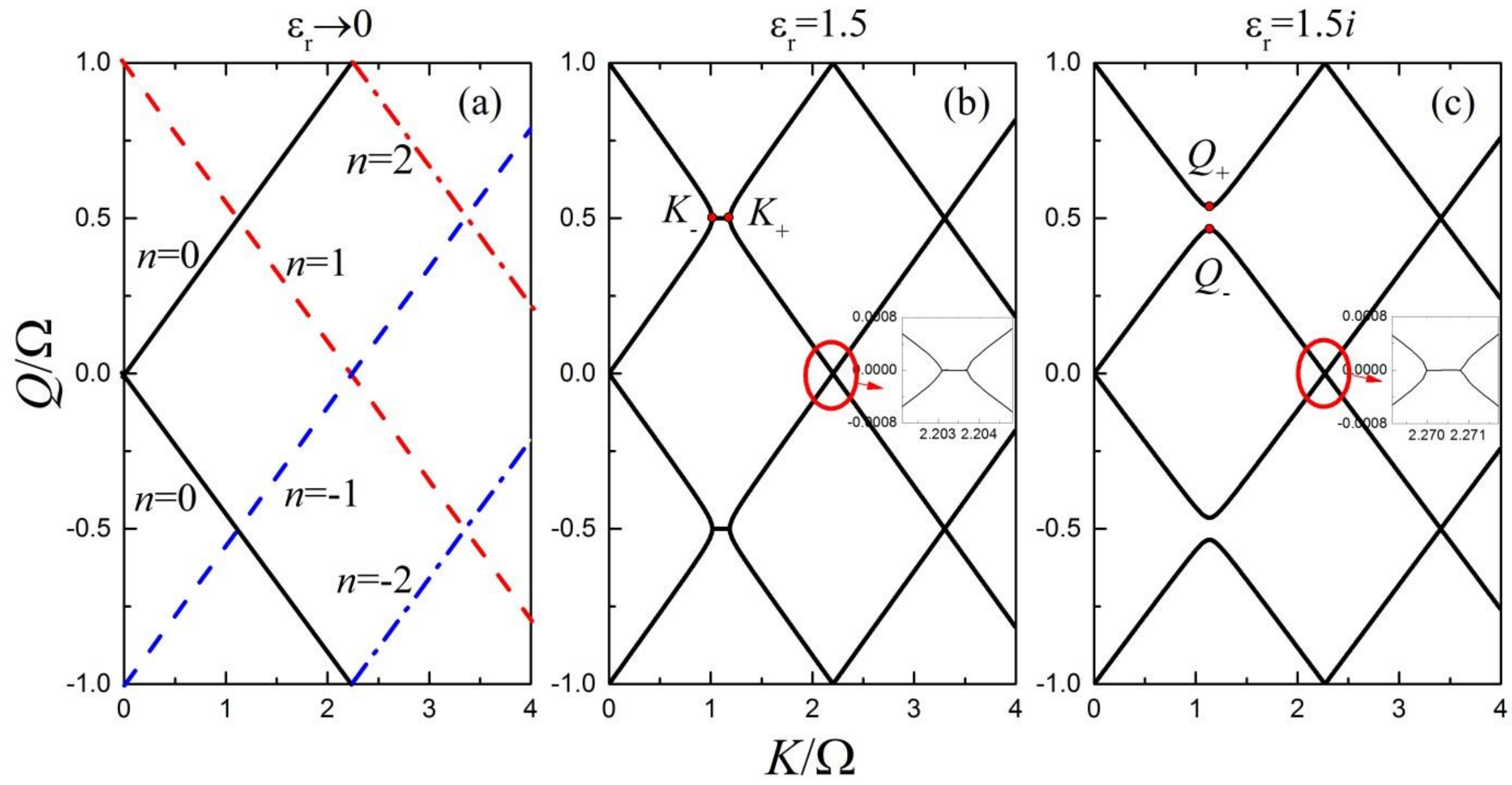

**Fig 1.** Floquet bands for the dynamic medium with permittivity given by $\varepsilon(t) = 5 + \varepsilon_r \sin(\Omega t + \phi)$. Only the real parts of the quasienergies are shown. (a) For $\varepsilon_r \to 0$, the bands of $\tilde{\mathrm{H}}_0$ are copied and shifted up and down by $n\Omega$ with $n$ being an integer. The crossings form degeneracy points. (b) For real $\varepsilon_r$, two bands in the vicinity of a degeneracy point will attract each other and form a pair of EPs. The Floquet states between the two EPs are in the broken phase. (c) When $\varepsilon_r$ is purely imaginary, two bands in the vicinity of a degeneracy point will repel each other and form a gap. The insets of Figs. (b) and (c) show the EPs formed by the coupling $\tilde{\mathrm{H}}_{\pm 2}$ for the cases of real and pure imagainary $\varepsilon_r$, respectively.

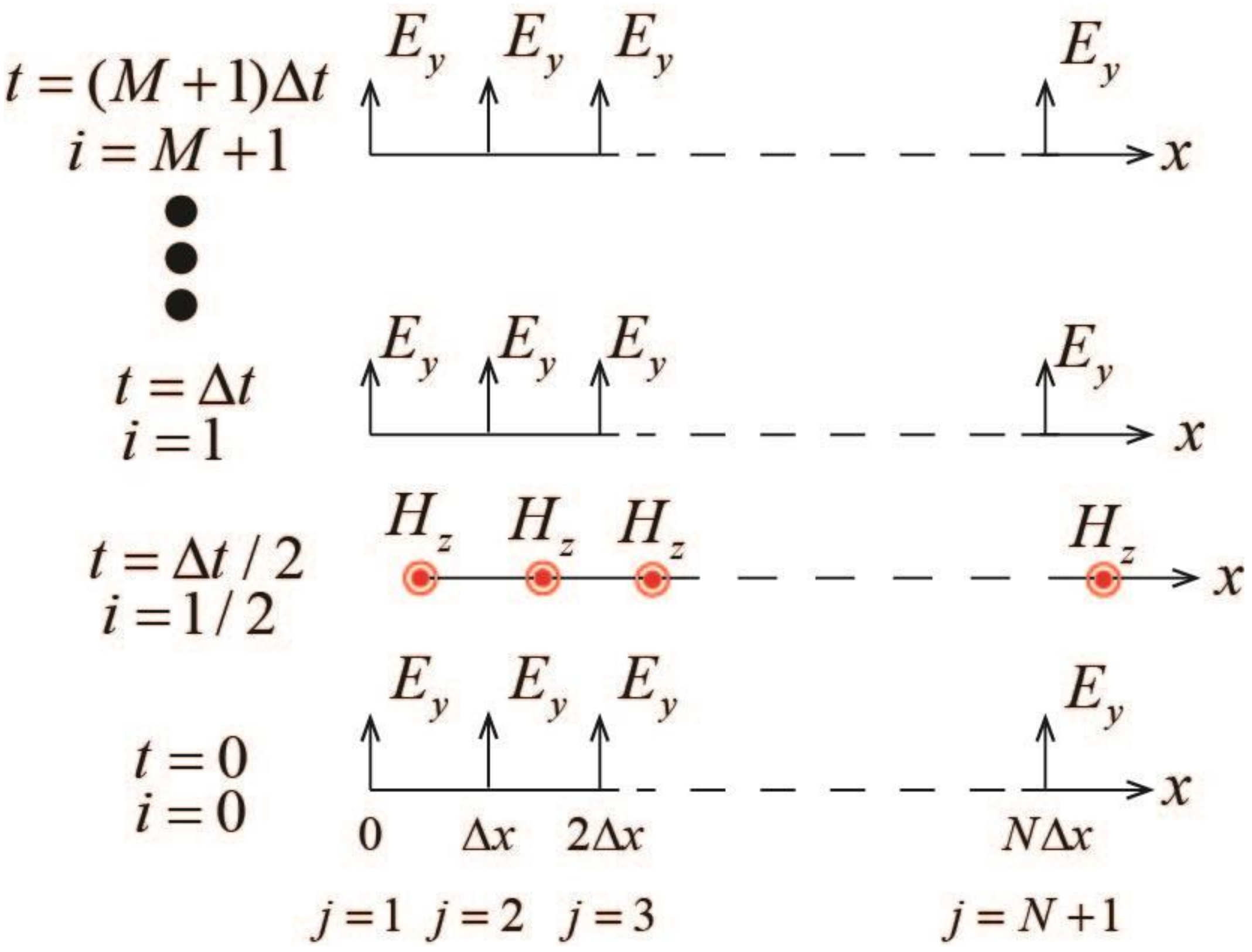

**Fig.2**. General representation of the finite-difference time domain scheme for numerical simulation of photonic Floquet media.

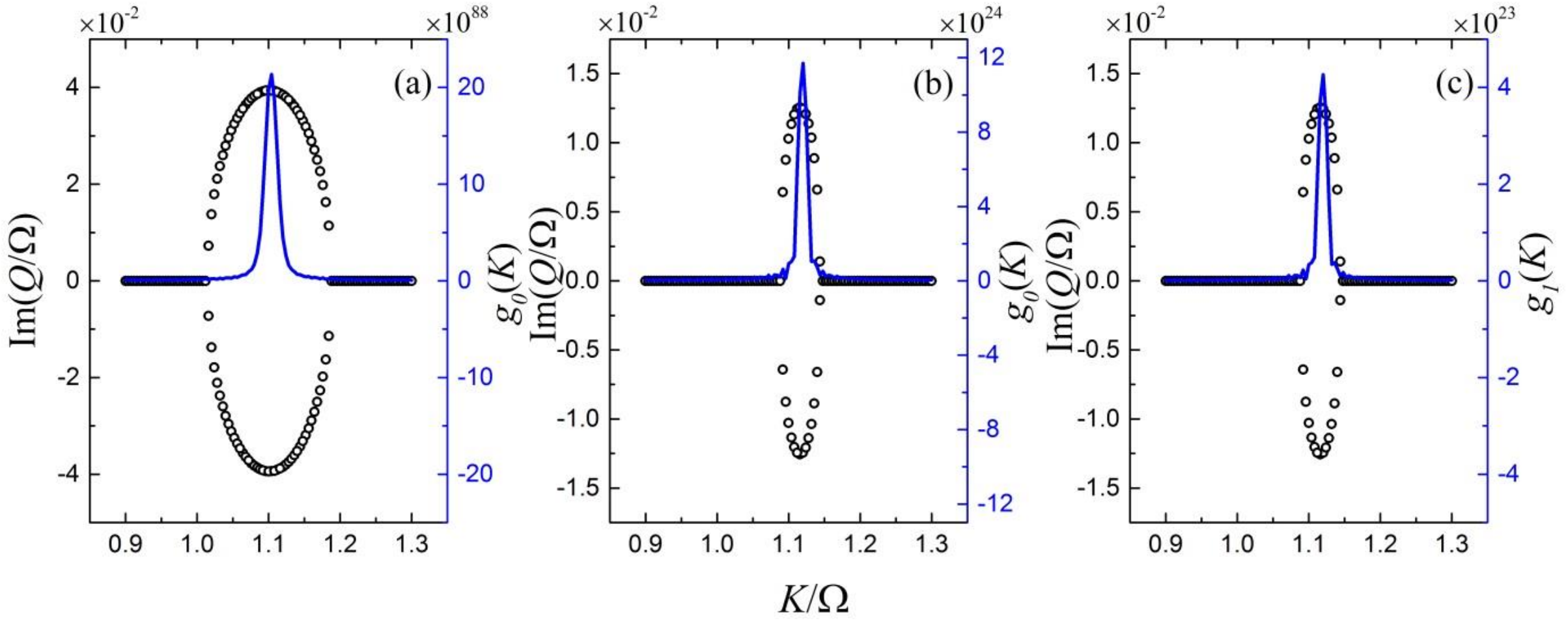

**Fig 3. (a)** Imaginary parts of the quasienergies (circles) and the absolute value of the function $g_0(K)$ (solid line) for $\varepsilon_r = 1.5$. (b) same as (a) but with $\varepsilon_r = 0.5$. (c) same as (b) but the function $g_1(K)$ is plotted. The functions $g_0(K)$ and $g_1(K)$ are defined in Eqs.(32) and (33) , respectively. For (a)-(c), the parameters used for the pulse are $t_0 = \tau = 250T$, and the starting and ending time for the integration of Eq. (32) are $t_1 = 410T, t_2 = 1010T$, where $T = 2\pi / \Omega$. The thickness of the slab is $L = 2000\pi c / \Omega$.

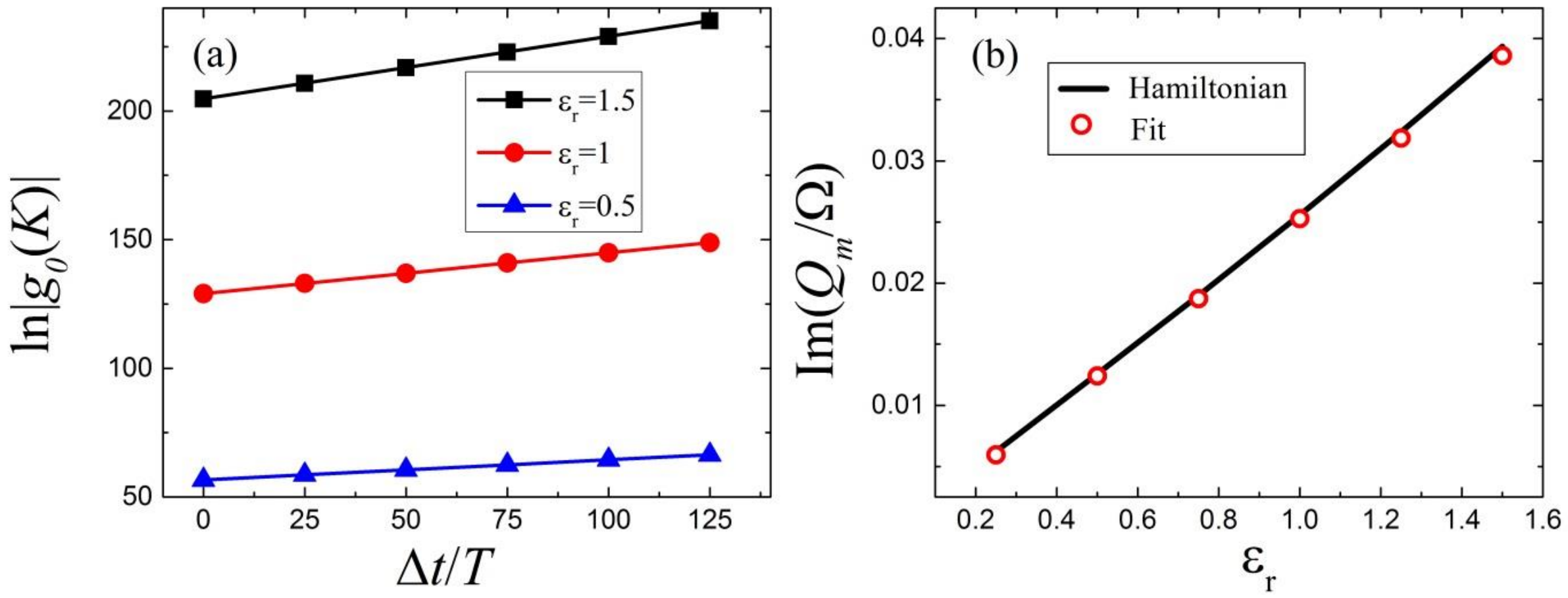

**Fig.4**. (a) The logarithm of the maximum $|g_0(K)|$ as function of the shifted time $\Delta t$. $T = 2\pi / \Omega$ is the modulation time period. (b) The maximum value of the imaginary quasienergy $Q_m$ as function of the modulation permittivity $\varepsilon_r$ calculated by the Floquet Hamiltonian (solid line) and the linear fitting of Fig. 4(a) slopes (red circles).

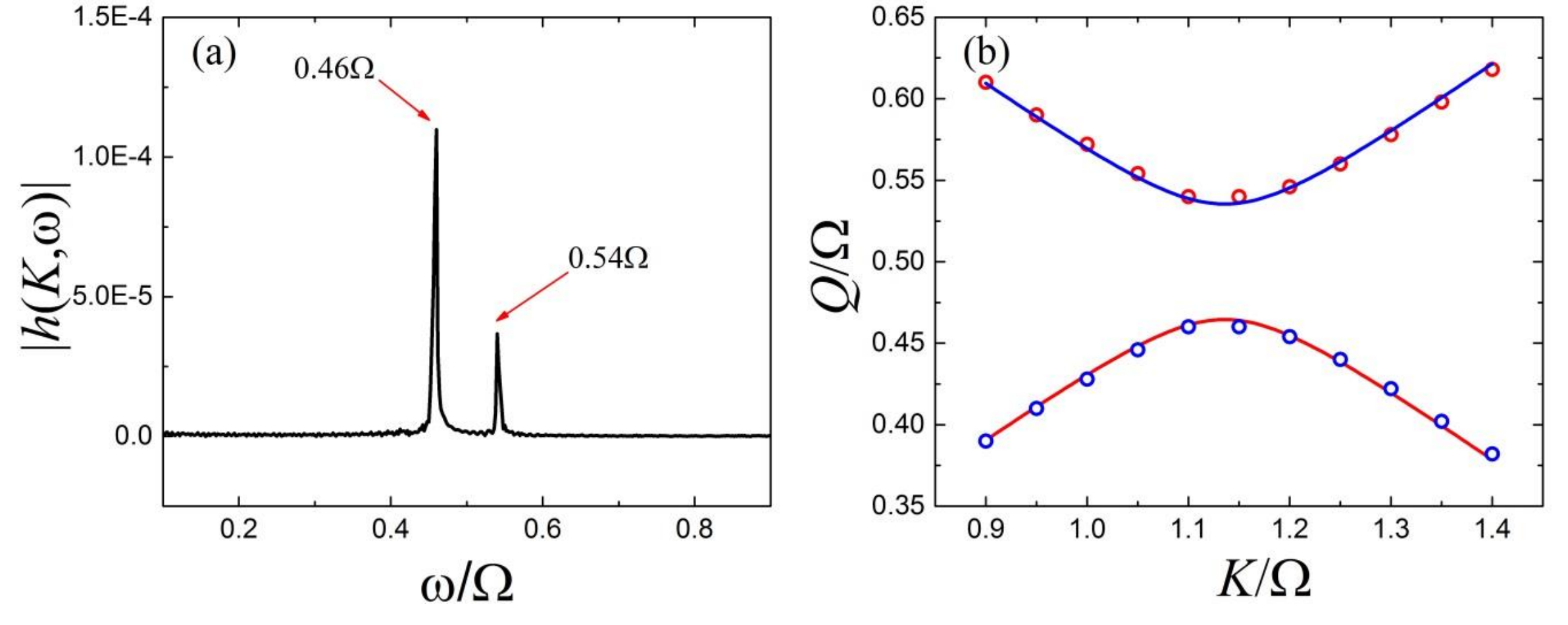

**Fig 5.** (a) The function $|h(K,\omega)|$ defined in Eq. (38) is plotted as a function of frequency $\omega$ for $K = 1.1\Omega$. Each peak corresponds to a mode. (b) The Floquet band structures in the range of $0.9\Omega \le K \le 1.4\Omega$. The solid lines are calculated using the Floquet Hamiltonian Eq. (20), and circles are determined by picking out the positions of the peaks of $|h(K,\omega)|$. The permittivity of the dynamic medium is $\varepsilon(t)=5+1.5\sin\Omega t$. The parameters for the pulse are $t_0=\tau=250T$, where $T=2\pi/\Omega$. For the integration of $h(K,\omega)$, we used $t_1=410T, t_2=1010T, L=2000\pi c/\Omega$.

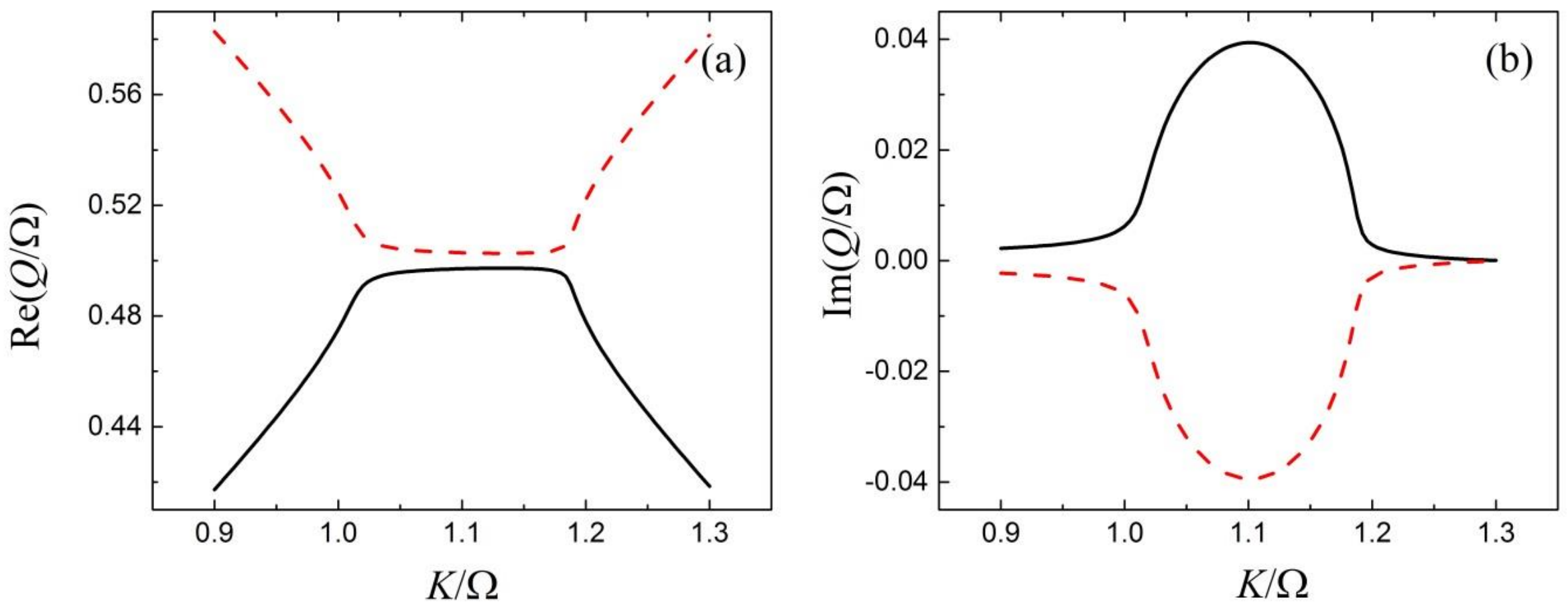

**Fig 6.** (a) The real part and (b) the imaginary part of the Floquet bands for the dynamic medium with time-dependent permittivity being $\varepsilon(t)=5+(1.5+0.1i)\sin(\Omega t+\phi)$. The black solid (red dashed) lines in (a) and (b) correspond to the same band.

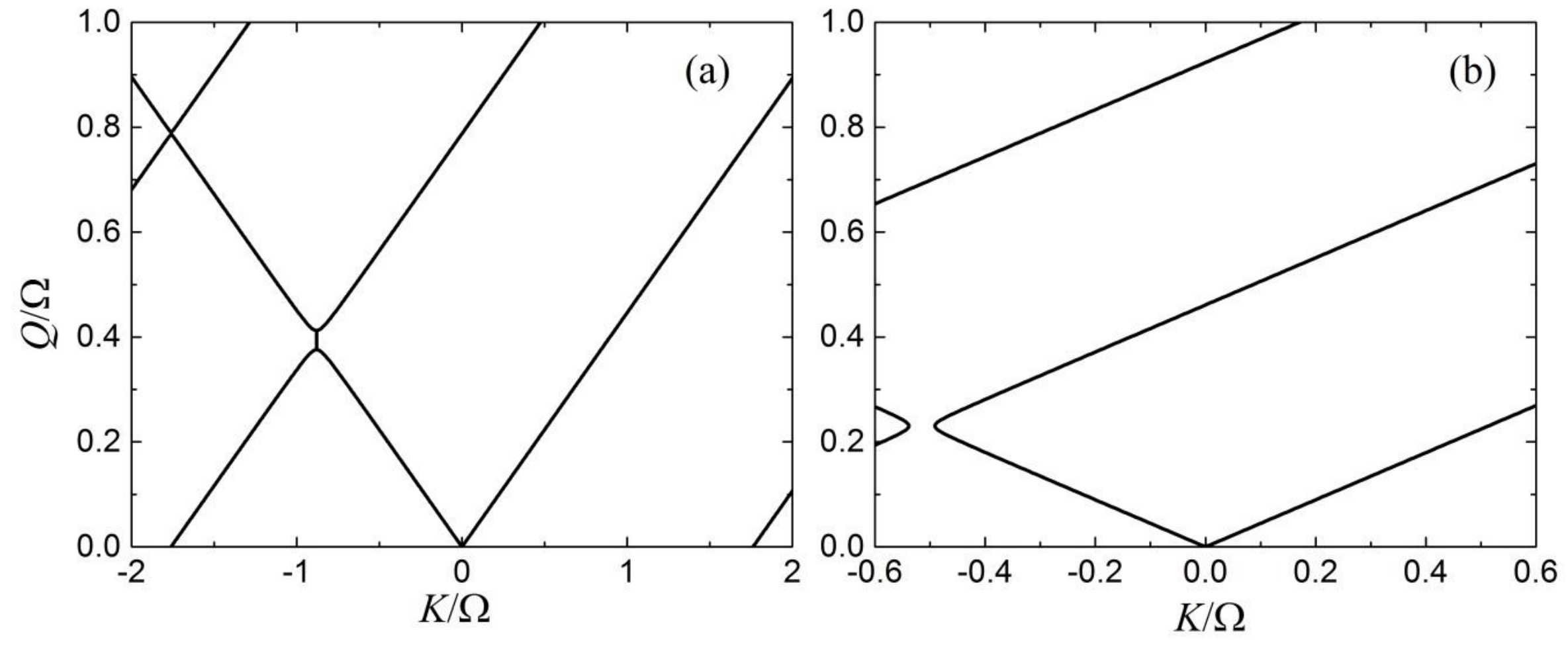

**Fig. 7**. Floquet bands for the space-time modulated medium with the spatial modulation frequency (a) $\beta=4\Omega$ and (b) $\beta=1.2\Omega$. The space-time modulated permittivity is given by $\varepsilon(x,t)=5+0.5\cos(\Omega t-\beta x+\phi)$.

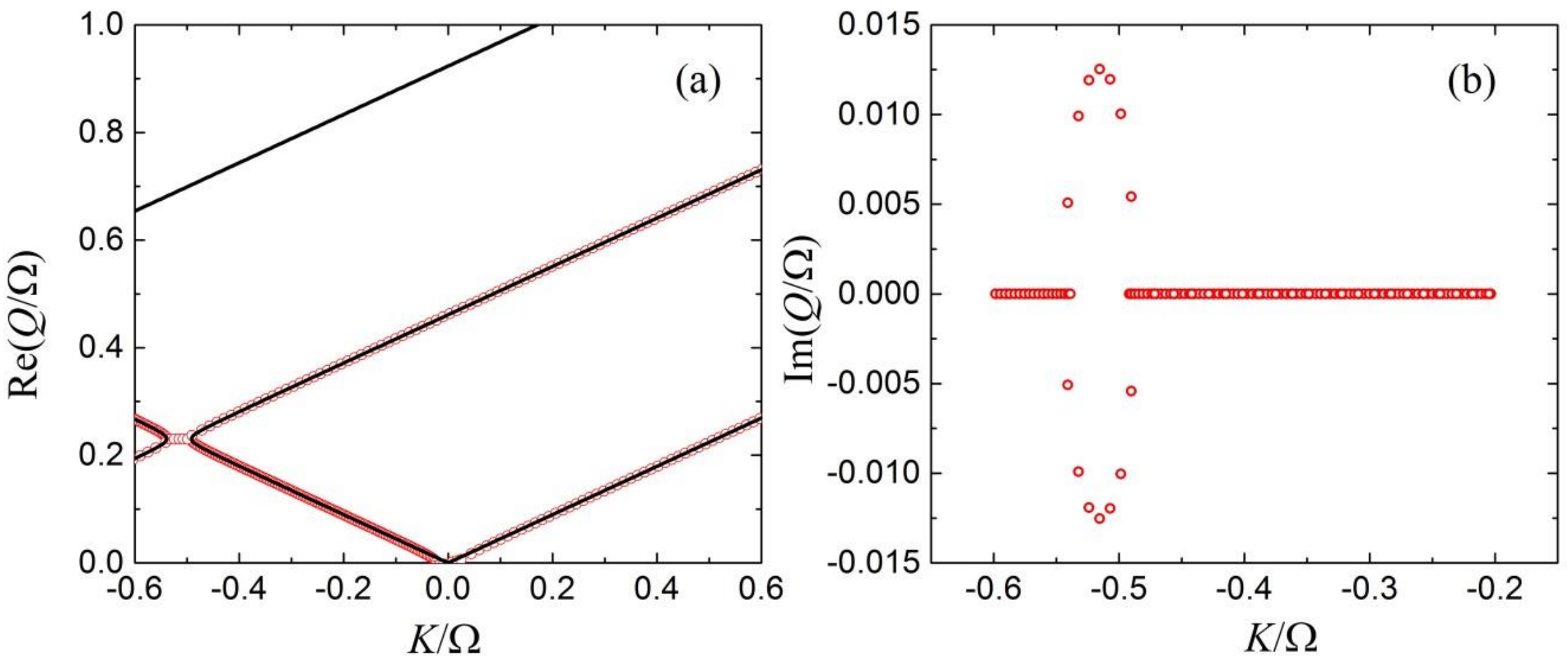

**Fig. 8**. (a) Comparison between the Floquet bands calculated by the plane wave expansion method (lines) and the Floquet Hamiltonian method (circles). (b) The imaginary part of the quasienergy as a function of wavenumber calculated using the Floquet Hamiltonian method.